# Optomechanical mass spectrometry

M. Sansa[a,1], M. Defoort[a,1,2], A. Brenac[b], M. Hermouet[a], L. Banniard[a], A. Fafin[a], M. Gely[a], C. Masselon[c,d], I. Favero[e], G. Jourdan[a], S. Hentz[a,3]

[a] Univ. Grenoble Alpes, CEA, LETI, 38000 Grenoble, France.

[b] Univ. Grenoble Alpes, CEA, CNRS, Grenoble INP, IRIG-Spintec, 38000 Grenoble, France.

[c] CEA, IRIG, Biologie à Grande Echelle, F-38054 Grenoble, France.

[d] Inserm, Unité 1038, F-38054 Grenoble, France.

[e] Matériaux et Phénomènes Quantiques, CNRS UMR 7162, Université de Paris, 75013 Paris, France.

[1] M. S. and M. D. contributed equally to this work.

[2] Present address: Univ. Grenoble Alpes, CNRS, Grenoble INP, TIMA, 38000 Grenoble, France

[3] To whom correspondence should be addressed. E-mail: sebastien.hentz@cea.fr

Abstract

It has been demonstrated in the recent years that nanomechanical mass spectrometry was well suited for the analysis of specific high mass species such as viruses. Still, the exclusive use of one-dimensional devices such as vibrating beams forces a trade-off between analysis time (related to capture area) and mass resolution (inversely proportional to device mass). Such devices also require complex readout schemes to simultaneously monitor multiple resonance modes, in particular when stiffness or shape come into play, which degrades mass resolution. These issues restrict nanomechanical MS to the analysis of species with specific properties. We report here the first demonstration of single-particle mass spectrometry with nano-optomechanical resonators fabricated with a Very Large Scale Integration process. The unique motion sensitivity of optomechanical techniques allows the use of resonators with new topologies. The device introduced here is designed to be impervious to particle position, stiffness or shape, opening the way to the analysis of large aspect ratio biological objects of great significance such as viruses with a tail or fibrils. Compared to top-down beam resonators with electrical read-out and state-of-the-art mass resolution, we show a three-fold improvement in capture area with no resolution degradation, despite the use of a single resonance mode. Short analysis time is in reach thanks to the sensitivity and bandwidth of optomechanical read-out and its compliance with telecom multiplexing techniques.

Main text

Introduction

Mass spectrometry (MS) is an extraordinary versatile analytical technique (1, 2) that relies on ionization, electromagnetic fields to manipulate ions, and ensemble averaging of ions' mass-to-charge ratios to characterize species based on their specific molecular mass. While MS was initially well suited for light objects such as molecules, it has been continuously making advances towards the characterization of ever larger supra-molecular assemblies (3–5). Yet, no commercially-available instrument can measure masses above a few MDa today.

An alternative approach consists in using nanomechanical resonators to determine the mass of individual particles accreting on their surface by measuring the induced frequency shift. Such nanomechanical mass sensing reached extremely low mass limits of detection (6). As early systems relied on ionization and ion guides however, they suffered from the same limitations as conventional MS, which limited operation to the low MDa range together with prohibitively long analysis times and excessive sample requirements due to the device's extremely low capture cross-section (7). Eventually, the realization that nanomechanical resonators were ideally suited for analysis of masses in the MDa to GDa range opened the way for MS of species irrespective of charge (8). Arrays of nanoresonators were demonstrated to circumvent the cross-section issue (9) and 100 MDa bacteriophage virus capsids were recently analyzed with a neutral nanomechanical MS system displaying high efficiency and resolution without the requirement for ionization (10).

The nanoresonators used in past mass experiments were all one-dimensional structures such as beams and cantilevers vibrating with flexural modes. With such devices, the frequency shift induced by a particle accreting on the resonator not only depends on mass, but it is also a function of landing position (7, 8). In some cases, particle stiffness (11), size and shape (12) have to be accounted for. This has deleterious consequences: complex read-out schemes must be used for simultaneous monitoring of multiple vibration modes; mass resolution dependence on landing position leads to a degradation of the average resolution by close to an order of magnitude compared to the optimum (10), forcing a drastic compromise on capture area. Most importantly, the use of such one-dimensional resonators impedes the analysis of many crucial particles of biomedical interest having large aspect ratio, such as tailed viruses or fibrils, in particular when their length is similar to or exceeds the resonator's width.

In parallel with the developments employing electrically-transduced nanoresonators, the field of cavity optomechanics has made critical advances (13, 14), approaching technological maturity. Its extreme displacement sensitivity and virtually unlimited bandwidth offers great promise for inertial (15, 16), chemical (17), biological (18), force (19) and mass (20) sensing. This is particularly the case for nanomechanical MS, with the prospect of smaller and more sensitive resonators operating at higher frequencies and the possibility to design new device topologies enabled by the optomechanical transduction.

Here we report the first demonstration of single-particle MS with optomechanical nanoresonators. These devices were fabricated with a unique Very Large Scale Integration process for optomechanics. Electrostatic actuation and released mechanical parts inherited from nano-electro-mechanical systems were used in combination with silicon photonics features such as on-chip light coupling and fiber packaging for vacuum and cryogenic operation. We took advantage of optomechanical transduction to measure the mass of single particles in real time with a device vibrating in a single resonance mode, featuring a capture area 3-fold larger than prior one-dimensional electrical resonators while maintaining similar mass resolution. This new device was designed to remain insensitive to particle landing position, stiffness, or aspect ratio, paving the way towards analyses of non-spherical particles.

Single-mode optomechanical nano-resonators

Our so-called "nano-ram" resonator consists in a sensing platform supported by four beams (Figure 1a). A most interesting feature of this device lies in its in-plane rigid body translation mode, in which the particle capture platform does not undergo deformation. The frequency shift induced by a particle on the platform thus remains independent from landing position (Figure 1b), particle stiffness, shape or aspect ratio (see Supplementary Figure 1). In this scheme, single-mode operation significantly simplifies readout and data processing, mass sensitivity remains constant and resolution optimal over the entire sensing area, while the larger capture area yields faster acquisition.

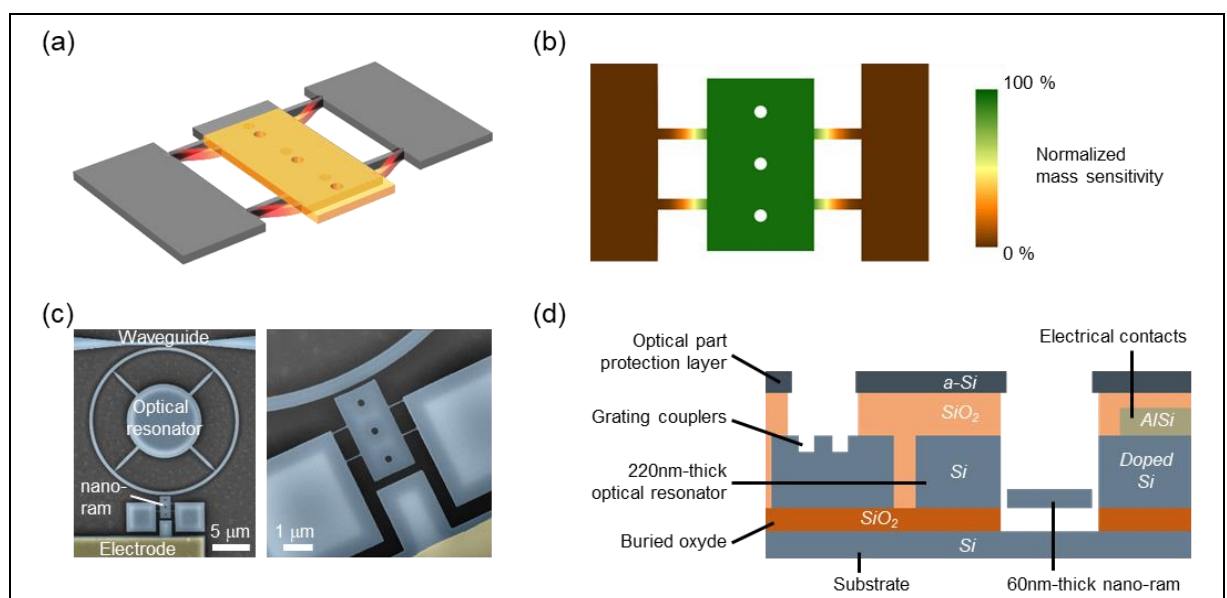

Figure 1 – **Single-mode optomechanical resonator for mass spectrometry**. a) In-plane rigid-body vibration mode of interest of the nanomechanical resonator (see Supplementary Figure 2 for the other modes). b) Finite-element color map of normalized frequency sensitivity to added point mass, showing that the frequency shift due to particle adsorption does not depend on particle position on the platform. c) False-colored scanning electron microscope images of the device, general view (left) and zoom-in on the nano-ram (right). The platform is 1.5 µm wide and 3 µm long, with $80 \times 500$ nm support beams. The optical ring diameter is 20 µm, and the optical ring-to-platform gap is 100 nm. Close to 1.55 µm wavelength light is coupled in and out of the ring by optical waveguides through a 200 nm gap. Electrostatic actuation is performed with a side-gate 250 nm away from the nanoresonator. (d) Cross-section showing the different components of the device. The silicon-on-insulator (SOI) top layer is 220 nm thick, partially etched to realize the optical grating couplers. The nanoresonator is etched down to 60 nm. The crystalline Si layer is highly doped locally for low metal-to-silicon contact resistance. A 200 nm amorphous silicon layer is deposited above a planarized silicon oxide layer for protection and etched open above the grating couplers and the nano-ram.

One of the main figures of merit of a resonator for mass sensing is the minimum detectable mass $\delta m_{\text{min}}$. It is given by

$$\delta m_{\text{min}} = 2 M_{\text{res}} \langle \frac{\delta f}{f_0} \rangle_{\text{min}} \qquad (1)$$

where $M_{\text{res}}$ is the total mass of the platform (see Supplementary Figure 2) and $\langle \frac{\delta f}{f_0} \rangle_{\text{min}}$ its frequency stability. Both factors should remain as small as possible to improve mass resolution. In order to maintain a large capture area while minimizing its total mass, the platform and its supports were thinned down to 60 nm. The frequency stability is ultimately limited by intrinsic fluctuations of the resonance frequency in the mechanical domain (21). Reaching this limit experimentally requires large signal-to-noise ratio (SNR), which is maximized when the resonator is driven up to the onset of mechanical non-linearity and when its thermomechanical noise dominates. Although this is achievable for bulky devices in in-plane rigid body motion with electrical transduction (22, 23), it becomes extremely challenging for orders-of-magnitude lighter, smaller and higher-frequency resonators with a few 10's nm thickness such as ours. It is nevertheless within reach of optomechanical transduction, thanks to the remarkable displacement sensitivity. Conversely, electrostatic actuation was needed to drive the resonator to high amplitudes with low voltage, eventually up to the onset of non-linearity. These dual requirements called for integration of both optical and electrical inputs/outputs in the same fabrication process. A scanning electron microscope image shows all elements of the device (Figure 1c), fabricated in an industrial-grade clean room; a schematic cross-section is shown Figure 1d (see Methods for a detailed description of the fabrication process).

The nanoresonator's motion detunes the optical ring cavity and, in the bad cavity limit where we operate (14), modulates the output optical power at the mechanical resonance frequency (Figure 2a, inset) with an amplitude $\Delta P$ per unit displacement: $\Delta P \propto Q_{\text{opt}} C_r g_{\text{om}}$ (15). The loaded optical quality factor $Q_{\text{opt}}$ depends on both intrinsic and extrinsic losses, which are of equal value at the so-called critical coupling, when the contrast $C_r$ is equal to 1 (24). We designed a single-mode, 10µm radius ring resonator (Supplementary note 1) coupled to a close-by waveguide (200 nm gap). This conservative design leads to operation in an under-coupling regime, with $C_r \sim 0.23$ and typical loaded quality factors in the order of $5 \times 10^4$ (Figure 2b and Supplementary Figure 3). The dominant loss mechanism was attributed to scattering due to interaction with the ring anchoring spokes (25), even though these were tapered at their extremities, down to ~100 nm. Setting the laser wavelength close to the point of maximum slope of the optical resonance, and demodulating the transmitted light power at a frequency close to the nanoresonator's resonance frequency, we measured the thermomechanical noise of the mechanical resonator (Figure 2c, inset). This allowed the determination of the optomechanical coupling factor $g_{\text{om}}$ (26), equal to 0.4 GHz/nm (Supplementary note 1). Light was coupled on-chip via grating couplers, designed so their maximum transmission was close to 1550 nm wavelength (27) (Figure 3b).

Figure 2a shows the measurement set-up, featuring optomechanical detection and electrical actuation. The nanoresonator's frequency responses were measured with increasing drive voltage (Figure 2c). Since actuation and motion sensing were not performed in the same physical domain, they were very well decoupled, ensuring a large signal-to-background ratio (Supplementary Figure 5). The frequency stability $\langle \frac{\delta f}{f_0} \rangle_{min}$ was measured by tracking the resonance frequency of the resonator in closed loop using a phase-locked-loop (PLL, Supplementary Figure 6a), and plotting the Allan deviation (28) for different drive voltages. Each plot was compared with open loop operation to make sure that the frequency stability was not affected by the PLL corrector (Supplementary Figure 6b). At low integration times and low drive voltages, *i.e.* in the regime where additive white noise dominates, the Allan deviation scaled like $\tau^{-1/2}$ and improved linearly with drive amplitude, as expected (21). Above some voltage threshold, the stability stopped improving and settled into a limit in the ppm range. As this threshold was well below the onset of non-linearity, operation in the non-linear regime was not required (29). This behavior was consistent with the presence of mechanical resonance frequency fluctuations (21). Despite the modest co-localization of mechanical motion and optical mode (due to the small thickness and lateral dimensions of the mechanical resonator), and despite small displacement and high frequency, the optomechanical transduction proved able to resolve thermomechanical noise (Figure 2c) and to reach the ultimate stability limit allowed by the mechanical device (Figure 2d). With a 1 ppm frequency stability and a 608 fg mass, Eq. 1 yields a 0.7 MDa (1.2 ag) minimum detectable mass. This was similar to the average mass resolution obtained with previously reported piezoresistive beam resonators (0.5 MDa (10)), albeit with a threefold increased capture area and a resonator topology immune to particle's shape, stiffness and landing position.

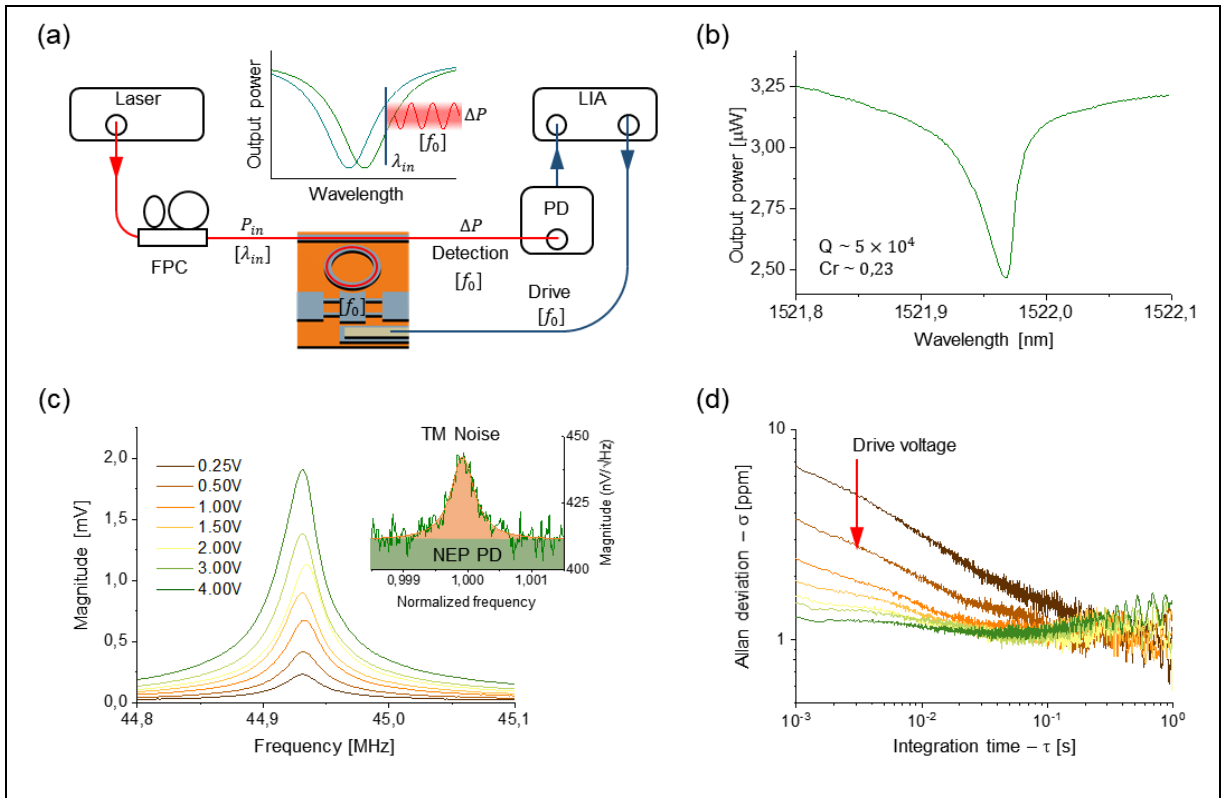

Figure 2 – **Optical, optomechanical and frequency stability measurements of the nanoresonator** at ~$10^{-5}$ Torr and 77 K. a) Optomechanical readout scheme. FPC stands for fiber-polarization controller, PD for photodetector, and LIA for lock-in amplifier. Red and dark blue lines are optical and electrical signals, respectively. The nanomechanical resonator vibrating at frequency $f_0$ modulates the resonance wavelength of the optical cavity and consequently the light output power at $f_0$. b) Narrow-band optical transmission spectrum around one resonance of the ring resonator measured with 0.5 mW output laser power. A slight thermo-optic behavior is observed. The loaded optical quality factor is ~$5 \times 10^4$, and the contrast is 0.23. c) Optomechanical spectra of the first in-plane mode, around 44.9 MHz with increasing drive voltages. The optical input power is set to 10 mW (Supplementary Figure 4). The mechanical quality factor is ~1700. Inset: thermomechanical noise of the resonator. The noise floor (green) is set by the photodetector's amplifier input noise, equivalent to a displacement resolution of $2.8 \times 10^{-14}$ m/$\sqrt{Hz}$. d) Allan deviation of the resonator for the same actuation voltages as panel (c), using a PLL. When increasing the drive voltage, the stability falls into a 1 ppm limit over the whole integration time range, which is consistent with the presence of mechanical resonance frequency fluctuations.

## Optomechanical mass spectrometry

In order to maintain a stable optical transduction during mass sensing experiments, the optical elements were protected with a ~200 nm-thick amorphous silicon layer (see cross-section Figure 1d) etched open only on top of the resonator's sensing platform (Figure 3d). This protective layer avoided degradation of both optical transmission and optomechanical coupling as well detuning of the optical cavity due to accretion of particles on the optical waveguide and resonator (Supplementary Figure 7). The alignment uncertainty between lithography levels was key in this context and sub-100 nm alignment was obtained (see Methods).

An additional challenge for routine mass sensing experiments was efficient light coupling on-chip within a spatially constrained particle-generating system under vacuum and at low temperature. Prior optomechanical works used tapered optical fibers directly coupled to the cavity (30) or microlensed fibers coupled to tapered waveguides (31) but both techniques require piezoelectric stages and proper vibration control. Light can also be focused with an objective lens onto grating couplers from outside the system (17) but this requires a device in immediate proximity and parallel to a window. In this work, we used fiber-transposer chips (Teem Photonics) that were UV-glued onto the grating couplers (Figures 3b, 3c and Supplementary Figure S8). This silicon photonics-derived technique allowed small footprint, high density (grating couplers are ~500 µm away from the device) and high transmission. Standard vacuum feedthroughs could thus be used to route both electrical and optical signals in and out of the system (Figure 3a).

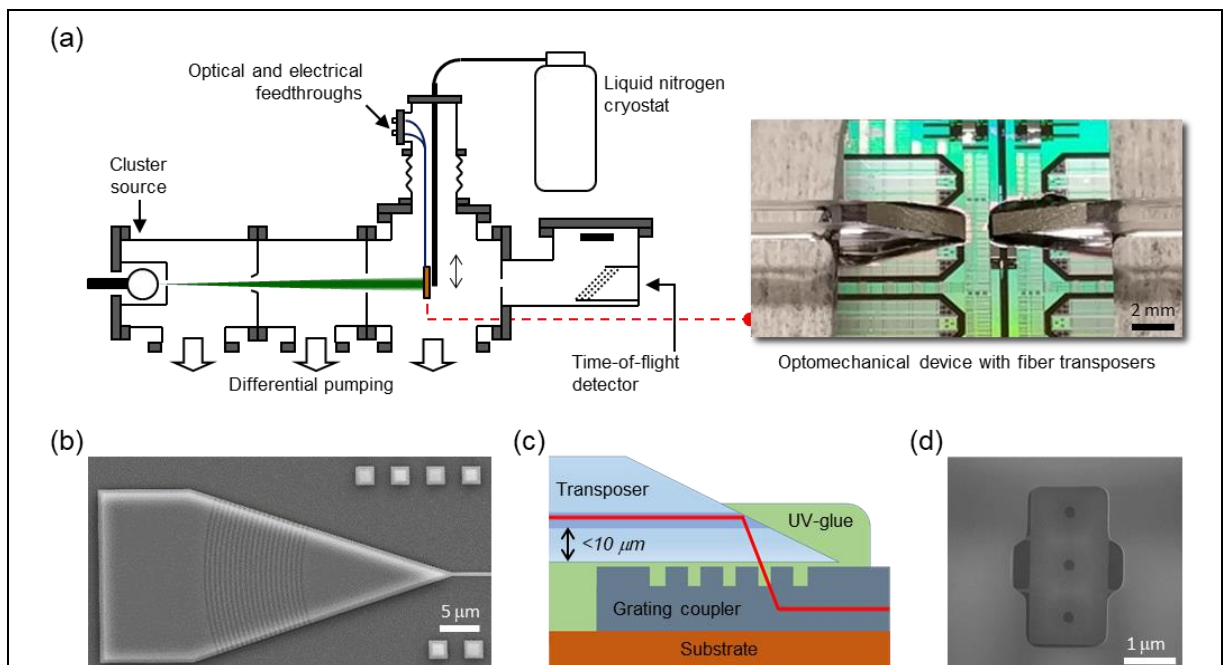

Figure 3 – **Vacuum system used for mass deposition and optically-packaged optomechanical device.** a) The set-up consists in a sputtering source capable of generating metallic clusters of controlled size and mass, and a Time-Of-Flight spectrometer. Vacuum feedthroughs carry electrical and optical signals in and out of the system. The optomechanical device is placed on a retractable sample holder at the end of a cryostat (liquid nitrogen). By moving the sample holder in and out of the cluster beam, a given cluster population can be measured with the nanoresonator and TOF detector sequentially. b) Light is coupled in and out of the optomechanical chips using grating couplers with a pitch of 0.6 μm and a width of 0.3 μm, designed for maximum transmission close to a 1550 nm wavelength and an input angle of 10° (27). c) Quasi in-plane optical packaging by waveguide-to-fiber transposer chips (measuring approximately $1 \times 2 \times 20$ mm), aligned and glued to the grating couplers. The electrical inputs/outputs are obtained by standard wire-bonding on metallic pads (see also Supplementary Figure 9). d) SEM image of the amorphous silicon protection layer covering all optical and electrical elements so that mass deposition only occurs on the sensing platform.

Mass spectrometry experiments were performed in a four-vacuum chambers system described in more detail elsewhere (32), that comprised a sputtering source and an in-line time-of-flight mass spectrometer (TOF-MS). Metallic nanoclusters beams with tunable size and deposition rate were generated using sputtering gas-aggregation technique. Nanoclusters were expelled through a differential pumping stage (Figure 3a) into a vacuum deposition chamber ($10^{-5}$ Torr) containing our optomechanical device mounted onto a translational stage that could be inserted into the cluster beam. Upon retraction of this stage, the particle flux entered the in-line TOF mass spectrometer, where the mass-to-charge distribution of charged particles could be determined. The configuration of the deposition chamber thus allowed TOF-MS and optomechanical MS measurements sequentially on the same cluster population.

We selected tantalum as analyte as it is both dense (16.6 g.cm$^{-3}$) and readily condenses into large clusters. Detecting such large clusters with the TOF mass spectrometer constituted a first challenge. We were able to detect clusters up to ~7 MDa by TOF-MS, *i.e.* close to two-fold heavier compared to previous experiments (8, 9). This was achieved by an increase in the ion acceleration voltage (from 3 to 3.6 kV), as the ion detector efficiency is proportional to the square of this parameter. Monitoring the nanoresonator's frequency was achieved with a PLL with a response time of ~10 ms. An example of mechanical frequency trace in closed-loop operation acquired during exposure to cluster beam is shown Figure 4a. Frequency jumps (see inset) were converted into mass with $\Delta m = 2 M_{\text{res}} \Delta f / f_0$. In contrast with prior devices, landing position, stiffness or aspect ratio did not have to be accounted for (see Supplementary Figure 1) and direct access to the mass information was possible. Tuning of the nanocluster source parameters yielded particle adsorption event rates on the order of four events per second, making the landing of several particles within the PLL response time very unlikely. Notably, we observed a three-fold improvement in event rate compared to one-dimensional vibrating beams with a similar particle flux (8, 9), which was in agreement with the capture surface ratio of the resonator used. Particle landing events were combined into histograms with bin widths ranging from 0.5 to 1 MDa. Mass spectra of four different nanocluster populations (2.8, 3.9, 5.7 and 7.7 MDa) acquired by both optomechanical MS and TOF-MS are displayed in Figure 4. While optomechanical MS directly provided the cluster mass distribution, TOF-MS yielded mass-to-charge ratio distributions. Although the vast majority of clusters carried only one charge, the presence of multiple charge states and the low TOF signal-to-noise ratio could make spectra interpretation complex (Supplementary Figure 12). Each optomechanical-MS spectrum was acquired in only 5 min, and comprised ~1000 events. Both techniques provided comparable mean mass estimates for low mass populations (2.76 vs. 2.83 MDa and 3.85 vs. 3.82 MDa for optomechanical MS and TOF-MS respectively). Nevertheless, the optomechanical spectra consistently displayed a larger proportion of higher mass particles than TOF-MS and this was even more so for high mass particles (5.67 and 7.68 MDa). This could be attributed to the fact that measurements were performed at the extreme mass range achievable for our TOF-MS: detection efficiency being roughly inversely proportional to ion mass, a fast decay of the detected signal with particle mass was expected, causing a distortion of the spectrum. This in turn degraded the detector's SNR, making baseline subtraction more difficult, and may have resulted in a high mass cut off, explaining the underestimation of the average mass and of the standard deviation. This was prominently observed with the population with the largest mass, for which obtaining a spectrum with TOF-MS proved very challenging and required long integration times. Conversely, the nanoresonator limit of detection remaining constant with mass, its resolving power (ratio of analyzed mass to mass resolution) increased with mass; and high-mass events were detected with better relative precision. This suggests that the high-mass tails of the optomechanical spectra were very likely representative of the cluster population, while TOF-MS "missed" these clusters (see Methods and Supplementary Figures 10-11). This most likely explains the discrepancy in mean mass for the two heavier cluster populations. On the other hand, the low mass events (typically below 3 MDa) were performed close to the resonator's limit of detection (~0.7 MDa). Measurement of the resonator's frequency noise was required to correctly interpret the associated spectra in this range (see Methods). In total, our optomechanical resonator accumulated close to 10% of its initial mass in less than an hour of deposition without measurable performance degradation (Supplementary Figure 13). The optical response remained constant during the whole process, without noticeable changes in quality factor, transmission or resonance wavelength detuning (Supplementary Figure 7).

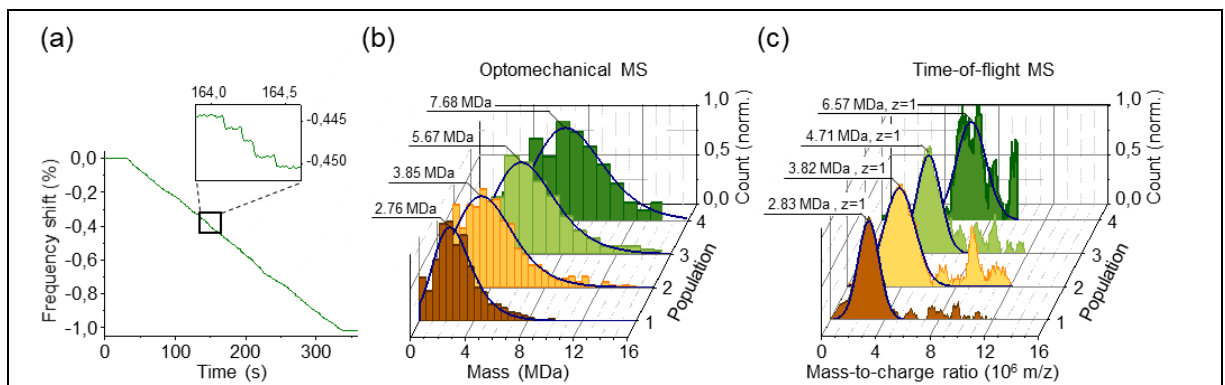

Figure 4 – **Single-particle optomechanical mass spectrometry of tantalum clusters.** a) Frequency trace of the optomechanical resonator for the light green (5.7 MDa) cluster population deposition. For this particular measurement, 1140 events were recorded in ~5 min, for a total deposited mass close to 1% of the mass of the resonator. The sampling time is here 50 µs, the PLL bandwidth is 10 ms and the plotted frequency trace is averaged to a 10 ms integration time. Spectra are plotted from detected frequency jump heights (Methods). Inset: detail showing several frequency jumps from individual cluster depositions. b) Normalized nanoresonator's and c) TOF mass spectra for 4 different tantalum cluster populations fitted with a log-normal function (dark blue lines) with mean masses ranging from 2.7 to 7.7 MDa (for optomechanical MS), equivalent to particle diameters from 8 to 11.3 nm. The number of events in the nanoresonator's spectra range from 534 to 1384, for measurement times up to 300 seconds in each case.

Outlook

Nanoresonators used in nanomechanical MS experiments to date were one-dimensional structures such as beams and cantilevers vibrating with flexural modes. With such structures, the frequency shift induced by accretion of a particle also depends on landing position: Measuring mass requires determining particle position on the beam. This is performed using a complex readout scheme designed to simultaneously monitor the frequency of at least two resonance modes (7, 8). In addition, the achievable mass resolution also depends on the landing position, and only close to half the total resonator's area is effectively used in the measurements (7–9). This in turns leads to reduced capture cross-section, longer analysis times and five to ten times larger average mass uncertainty compared to the best performance attainable on the beam (10). Moreover, measuring stiffness, size or shape may be required to correctly determine mass, which means monitoring an even higher number of modes (11, 12). Lastly, measuring the mass of elongated objects, if possible at all, would produce significant mass errors, in particular when their length is larger than the beam's lateral dimension.

The present report constitutes a significant departure from previous work as we successfully performed single-particle mass spectrometry with a resonator that requires monitoring a single vibration mode and a simple and efficient readout scheme. This resonator is by design insensitive to landing position, particle stiffness, size, shape or aspect ratio. It embeds a platform that vibrates in-plane with a rigid-body motion, providing large capture cross-section. The sensing platform was thinned down in order to obtain low resonator mass and high frequency-to-mass sensitivity. Obtaining excellent mass resolution requires large signal-to-noise ratio, large dynamic range and high displacement sensitivity. This is extremely difficult to achieve with electrical transductions for such thin, in-plane devices and we used here on-chip cavity-based optomechanical resonators for this purpose. They were fabricated with the first Very Large Scale Integration fabrication process for optomechanics, on 200 mm wafers. Using this device, we have performed MS analysis of tantalum nanoclusters ranging from 2.8 to 7.7 MDa in less than 5 minutes, and demonstrated an excellent stability of the sensor during these depositions.

This first demonstration of on-chip optomechanics as a superior alternative to electromechanical resonator for high-resolution single-particle mass spectrometry paves the way to high-throughput MS analysis of synthetic and natural nanoparticles with any possible geometry. For this purpose, our photonics-derived fabrication process is easily amenable to multiplexing of a large number of resonators using standard wavelength-division multiplexing (33) and packaging techniques. In addition, while our resonators were etched down to 60 nm, this thickness could be reasonably reduced to ~10 nm in the future. Mass resolution below 100 kDa could be reached with such lighter resonators and much stiffer anchors, yielding resonance frequency above 500MHz, still in reach of optomechanical read-out (34). Such devices would open new applications in structural biology and nano-characterization by achieving the performance required for the analysis of non-spherical objects that are challenging for current nanomechanical MS, such as amyloid fibrils involved in neurodegenerative diseases, or tailed viruses used for phagotherapy, a promising alternative to standard antibiotherapy.

## Methods

### Fabrication process.

The sensors were fabricated with a Very Large Scale Integration fabrication process in an industrial-grade clean-room. Starting from 200 mm Silicon-on-Insulator wafers with 220 nm thick silicon top layer, a first lithography and partial etching (70 nm) of the grating couplers are performed, followed by a second similar step to reduce the thickness of the nano-ram (the final thickness is measured by both Focused Ion Beam and spectroscopic ellipsometry). Next, the waveguides, optical cavities and mechanical resonators are patterned by dry etching. We developed a specific lithography step by variable shape beam (VSB) for this step. This technique allows both smooth pattern walls and low optical losses, and full electron beam resist insulation over the whole wafer in a reasonable amount of time (a few hours). Silicon is locally highly doped (boron, $5.10^{19}$ at/cm$^3$) for electrical actuation while preserving good optical properties elsewhere. A metal layer is deposited and patterned for electrical contacts. Then a sacrificial layer (silicon oxide) is deposited everywhere and planarized, followed by a deposition of a ~200 nm-thick amorphous silicon layer, which is etched open only on top of the resonator's sensing platform. A good alignment (below 100 nm) is critical for this step, as the aperture has to be located precisely on top of the platform. Finally the nanoresonator is released by vapor HF etching. Figure 1(d) shows a simplified cross-section of the device.

### Frequency data processing for spectrum plot.

The frequency time traces are scanned for abrupt changes to detect particle landing events. Because frequency traces are sampled, the jumps may spread over several sampling times. Therefore, their heights are quantified by measuring frequency before and after these abrupt changes and masses of the particles are inferred from those spread jumps. Masses are then binned to plot the histograms. In addition, frequency noise may produce false positives when dealing with masses close to the nanoresonator's resolution. We perform a statistical substraction of these events by measuring the nanoresonator's resonance frequency for 30 s before and after each deposition run. "Mass" histograms are built with these sets of frequency noise and their height is subsequently scaled as a function of the deposition run duration. Each bin value of the noise histograms is then subtracted from the deposited mass histogram.

Acknowledgements

Funding: We acknowledge support from the European Union through the ERC Enlightened project (616251) and the Marie-Curie Eurotalents incoming fellowship (M.S.).

We thank Florent Gardillou (Teem Photonics) for his help with the optical packaging.